\documentclass[amsthm,a4]{PoS}

\usepackage{bm}
\usepackage{graphicx}
\usepackage{amsmath}
\usepackage{amsthm}
\usepackage{amsfonts}
\usepackage{amssymb}
\newcommand{\tht}{\textheight}
\newcommand{\ig}{\includegraphics}

\title{Two-particle Correlation Functions with Distilled Propagators}

\ShortTitle{two-particle distillation}

\author{J.~Bulava\\
        Carnegie Mellon University\\
        E-mail: \email{jbulava@cmu.edu}}
\author{\speaker{K.~J.~Juge}\\
        University of the Pacific\\
        E-mail: \email{kjuge@pacific.edu}}
\author{C.~J.~Morningstar\\
        Carnegie Mellon University\\
        E-mail: \email{colin\_morningstar@cmu.edu}}
\author{M.~J.~Peardon\\
        Trinity College Dublin\\
        E-mail: \email{mjp@maths.tcd.ie}}
\author{C.~H.~Wong\\
        Carnegie Mellon University\\
        E-mail: \email{chikhimw@cmu.edu}\\\\
\centerline{{\bf For the Hadron Spectrum Collaboration} }}

\abstract{Correlation functions of the simplest multi-particle state will be presented using distilled quark propagators. The $I=2$ $\pi\pi$ state can be simulated without computing disconnected diagrams and thus is the simplest two-particle state that can be studied with quark sources placed on a single time-slice alone. We study the quality of the signals of this $\pi\pi$ correlation function using the quark-smearing guided distillation method. Results will be presented for $\pi\pi$ correlation functions computed on dynamical, anisotropic lattices.}

\FullConference{The XXVII International Symposium on Lattice Field Theory - LAT2009\\
		 July 26-31 2009\\
		 Peking University, Beijing, China}

\begin{document}

\section{Introduction}

Hadron spectroscopy (of excited states) on configurations with light dynamical quarks is reaching a stage where one needs to confront the issue of the mixing of single particle states with multi-particle states. If the goal is to determine the entire low-lying hadron spectra, one must be able to disentagle the single particle states from the multi-particle channels as the thresholds are approached with smaller quark masses. 

One of the key issues, then, is the simulation of multi-particle states with controled statistical errors. Simulating the excited state spectra without the use of explicit multi-particle operators seem to miss some of the excited states due to a poor overlap with the state. One could construct multi-particle operators with point-to-all quark propagators, but the need for various momenta operators at the source and sink prove to be expensive as is the construction of extended hadron operators. 
All-to-all quark propagators are too expensive for both computational and storage requirements. There are various ways of stochastically estimating all-to-all quark propagators in order to overcome this problem. In this paper, we test a single-timeslice to all-timeslices propagator method which is exact up to an ultraviolet cutoff. 

The simplest two-particle state which can be simulated with this method is the isospin-2 $\pi\pi$ channel since there are no disconnected diagrams in this channel (see Refs.~\cite{JLQCD}-\cite{Beane:2007xs} for recent dynamical simulations). The method we use is the distillation method (Ref.~\cite{Peardon:2009gh}), proposed by the Hadron Spectrum Collaboration. The possibility of computing scattering lengths and phase shifts via the finite volume method in Euclidean space (Ref.~\cite{Luscher:1986pf}) will also be discussed as preliminary results are presented. 

\section{Construction of Operators/Correlators}
The distillation process is described in detail in an earlier paper. We briefly summarize the method here for pion and two-pion correlation functions. 
\subsection{Distillation}
Hadron correlation functions are usually constructed from quark propagators which have been smeared in some way to reduce the high frequency noise from the signal. Jacobi smearing is a commonly used smearing method which amounts to a Gaussian-like damping of the high energy modes:
$$
\tilde{\psi}(\vec{x},t)=J_{\sigma,n_\sigma}(t)\psi(\vec{x},t)=\left(1+\frac{\sigma}{n_\sigma}\nabla^2(t)\right)^{n_\sigma}\psi(\vec{x},t)
$$
where $\nabla^2$ is the lattice Laplacian operator. Two parameters are tuned to increase the overlap of the operator onto the low-energy sector of the particular channel of interest, $\sigma$ and $n_\sigma$. The distillation operator on a particular timeslice is constructed from the eigenvectors, $v_x^{(k)}(t)$, of the Laplacian on that timeslice,
\begin{align}
\Box_{xy}(t)\equiv&\sum_{k=1}^nv_x^{(k)}(t)v_y^{(k)\dagger}(t) \\
\equiv&\mathbf{V}(t)\mathbf{V}^\dagger(t)
\end{align}
where $n$ is a number between $1$ and $M=N_cN_xN_yN_z$. We have used $n=64$ in this study. We then apply this distillation operator on our quark fields which are then used to construct the hadron interpolating operators. 
\subsubsection{Single Pion Correlation Function}
The correlation function of a single pion operator with arbitrary momentum is given by the standard expression,
$$
C_\pi(t,t_0)=\langle\overline{\tilde{\psi}}\gamma_5\psi(\vec{x},t)\overline{\psi}\gamma_5\psi(\vec{x}_0,t_0)\rangle.
$$
In order to simulate pions with definite momenta, we need to sum over the spatial sites at the source and the sink with the appropriate phases,
$$
C_\pi(\vec{p};t,t_0)=\langle\sum_{\vec{x}}e^{-i\vec{p}\cdot\vec{x}}\overline{\psi}\gamma_5\psi(\vec{x},t)\sum_{\vec{x}_0}e^{i\vec{p}\cdot\vec{x}_0}\overline{\psi}\gamma_5\psi(\vec{x}_0,t_0)\rangle.
$$
The sum over the spatial sites makes it necessary to construct the hadron interpolating $\overline{\psi}\gamma_5\psi$ operator on all spatial points. This is readily done at the sink even with point-to-all propagators, but requires substantial more work at the source. 
The distilled pion operator, on the other hand, requires no extra work as it contains the propagator from one timeslice to all other timeslices. The distilled correlation function is then given by,
\begin{align}
C_\pi(\vec{p};t,t_0)=&\langle\sum_{x,y,z}e^{-ipy}\overline{d}_x(t)\Box_{xy}(t)\gamma_5\Box_{yz}(t)u_z(t)\sum_{x_0,y_0,z_0}e^{ipy_0}\overline{u}_{x_0}(t_0)\Box_{x_0y_0}(t_0)\gamma_5\Box_{y_0z_0}(t_0)d_{z_0}(t_0)\rangle\\
=&\sum_{x,y,z}\sum_{x_0,y_0,z_0}\langle\Box_{y_0z_0}(t_0)d_{z_0}(t_0)\overline{d}_x(t)\Box_{xy}(t)\gamma_5\Box_{yz}(t)u_z(t)\overline{u}_{x_0}(t_0)\Box_{x_0y_0}(t_0)\gamma_5\rangle\\
=&\sum_{x,y,z}\sum_{x_0,y_0,z_0}\langle v^{(k)}_{y_0}(t_0)v^{(k)\dagger}_{z_0}(t_0)d_{z_0}(t_0)\overline{d}_x(t)\gamma_5v^{(j)}_{x}(t)v^{(j)\dagger}_{y}(t)v^{(l)}_{y}(t)v^{(l)\dagger}_{z}(t)\nonumber\\
&\ \ \ \ \ \ \ \ \ \ \ \ \ \ \ \ \ \ \ \ u_z(t)\overline{u}_{x_0}(t_0)v^{(i)}_{x_0}(t_0)v^{(i)\dagger}_{y_0}(t_0)\gamma_5\rangle\\
=&4\left[\sum_{x,z_0}v^{(j)\dagger}_{x}(t)M^{-1}(x,z_0) v^{(k)}_{z_0}(t_0)\right]^\dagger\left[\sum_ye^{-ipy}v^{(j)\dagger}_{y}(t)v^{(l)}_{y}(t)\right]\nonumber\\ 
&\ \ \ \ \ \ \ \ \ \ \ \ \  \left[\sum_{x_0,z}v^{(l)\dagger}_{z}(t)M^{-1}(z,x_0)v^{(i)}_{x_0}(t_0)\right]\left[\sum_{y_0}e^{ipy_0}v^{(i)\dagger}_{y_0}(t_0)v^{(k)}_{y_0}(t_0)\right]
\end{align}
The eigenvectors and perambulators have been computed and stored for the meson/baryon study (Ref.~\cite{HueyWen}). 
\subsection{Two Pion Correlation Function}
The two pion correlation function in the isospin-2 channel consists of only two diagrams; one of which involves a ``quark exchange'' (`C') diagram and the other which is simply the product of two, single pion correlation functions ($C_\pi(\vec{p};t,t_0)$) with back-to-back momenta (`D'). The quark exchange diagram is formed by contracting the perambulators in the following way:
\begin{align}
C^{cross}_{\pi\pi}=&4\left[v^{(j)\dagger}_xM^{-1}(x,z_0)v^{(k)}_{z_0}\right]^\dagger\left[e^{-ip\cdot y}v^{(j)\dagger}_yv^{(l)}_y\right]\left[v^{(l)\dagger}_zM^{-1}(z,x_0)v^{(i)}_{x_0}\right]\left[e^{-ip\cdot y_0}v^{(i)\dagger}_{y_0}v^{(k)}_{y_0}\right]\nonumber\\
&\times\left[v^{(r)\dagger}_wM^{-1}(w,w_0)v^{(l)}_{w_0}\right]^\dagger\left[e^{ip\cdot q}v^{(r)\dagger}_qv^{(s)}_q\right]\left[v^{(s)\dagger}_uM^{-1}(u,u_0)v^{(h)}_{u_0}\right]\left[e^{ip\cdot q_0}v^{(h)\dagger}_{q_0}v^{(l)}_{q_0}\right]
\end{align}
where repeated indices are summed over. The two pions have back-to-back momenta so that the total momentum of the two-particle system is zero. The two pion correlation function is formed by taking the difference of the two different contractions to project out the $I=2$ channel,
$$
C_{\pi\pi}(t,t_0)=\sum_{\vec{p}}\left[C_\pi(\vec{p};t,t_0)C_\pi(-\vec{p};t,t_0)-C^{cross}_{\pi\pi}(\vec{p},-\vec{p};t,t_0)\right]
$$
and the sum over momenta is done to project out the $s$-wave scattering state. 

\section{Simulation/Results}
\subsection{Parameters}
We use anisotropic, $2+1$ dynamical lattices whose tuning of the parameters are discussed in Ref.~\cite{Edwards:2008ja}. We have results from two different volumes ($16^3\times128$ and $20^3\times128$) with $m_\pi L\approx 3.5$ and $m_\pi L\approx 4.8$ (Ref.~\cite{Lin:2008pr}). The pion mass is roughly $360\ \rm{MeV}$ on both lattices. The lattice spacing is such that $r_0/a_s=3.221(25)$ with the renormalized anisotropy tuned to $\xi=a_s/a_t=3.5$. We use the lowest 64 eigenvectors of the Laplacian operator to construct the distilled propagators. The number of configurations analyzed were $100$ and $94$ for the $16^3\times128$, $20^3\times128$ volumes, respectively. 

We have used pion operators with momenta $0,\ 1,\ \sqrt{2},\ \sqrt{3}$ and $2$ (in spatial lattice units). All of the cross correlations were measured to obtain the full matrix of correlation functions.

\subsection{Analysis}
The five-by-five matrix of correlation functions was diagonalized to get the energies of the first five $\pi\pi$ scattering states. There are several ways of performing the diagonalization, each with its own set of advantages and disadvantages. The main concern is the contamination of the signal in each of the levels from higher lying states. We provide here some evidence for stability of our data against the different methods of diagonalization. 
The general method of extracting excited state energies from a matrix of correlation functions has been laid out in Ref.~\cite{Campbell:1987nv,Luscher:1990ck}. One starts with the correlation matrix $C_{ij}$,
$$
C_{ij}(t)=\langle{\mathcal O}_i(t){\mathcal O}_j^\dagger(0)\rangle
$$
and solve the generalized eigenvalue problem,
$$
C_{ij}(t^*)w_j=\lambda(t^*,t_0)C_{ij}(t_0)w_j
$$
for a given choice of $t_0$ and $t^*$ to compute the optimized, fixed-coefficient correlation functions. One can then compute the effective mass for each choice and also fit the optimized correlation functions to extract the various energy levels. This method does not guarantee that the excited state contamination is from states higher than the $N^{th}$ level, but the extraction of the levels does not involve any more manipulation of the data and can be fit using a simple, correlated chi-squared fit.

\begin{figure}[t]
\begin{center}
\begin{tabular}{cc}
\begin{minipage}{73mm}
\begin{center}
\ig[height=0.33\tht,width=0.28\tht,angle=270]{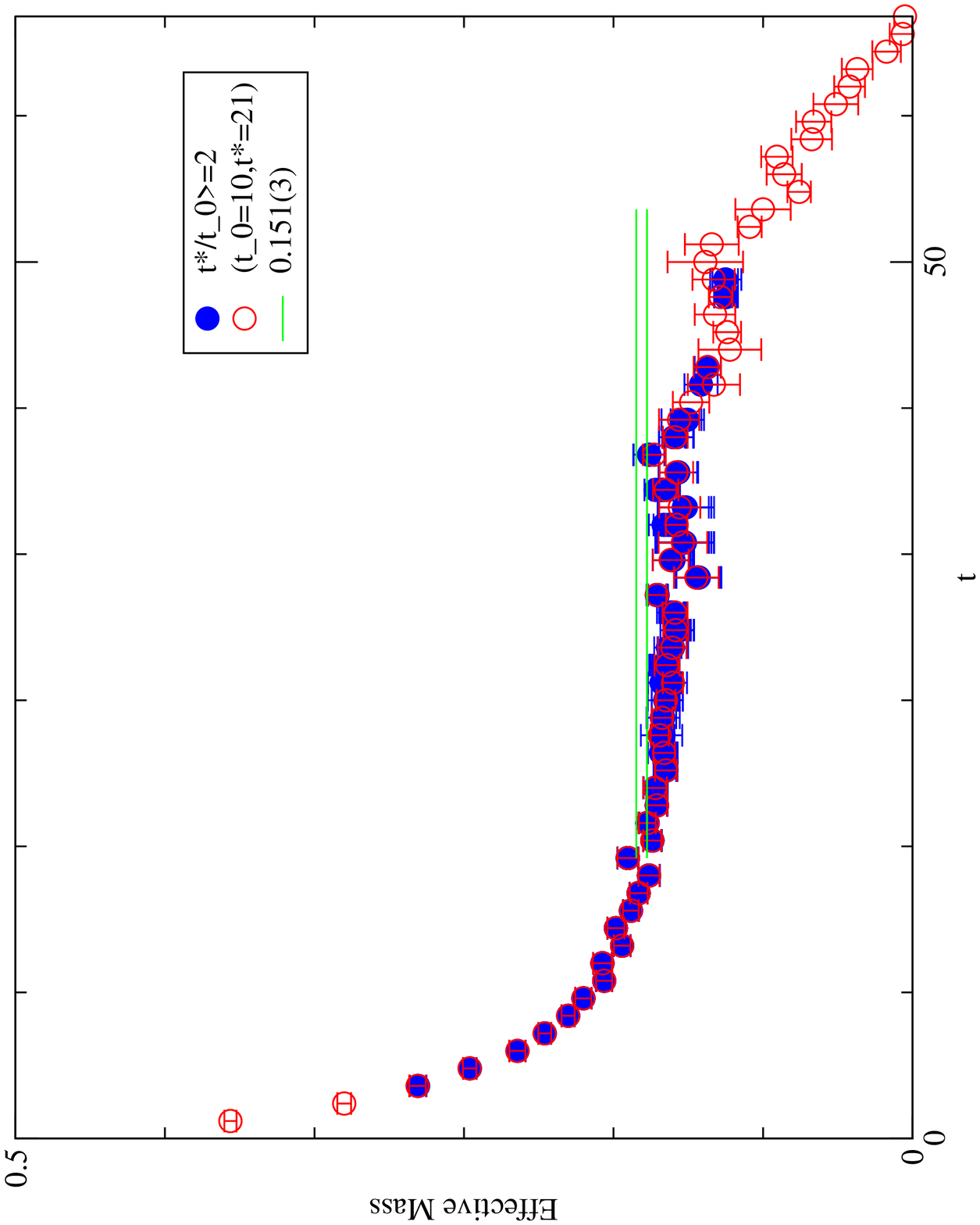} 
\caption{The optimized distilled pi-pi effective mass for the {\bf ground state} (${\bf t_0=10,t^*=21}$) on the $16^3\times128$ lattice with 64 eigenvectors. A single time-slice was used for the source operator and the standard definition of the effective mass was used with $\Delta t=1$.}
\label{fig:opt0_10}
\end{center}
\end{minipage}&
\begin{minipage}{73mm}
\vspace{4mm}
\begin{center}
\ig[height=0.33\tht,width=0.28\tht,angle=270]{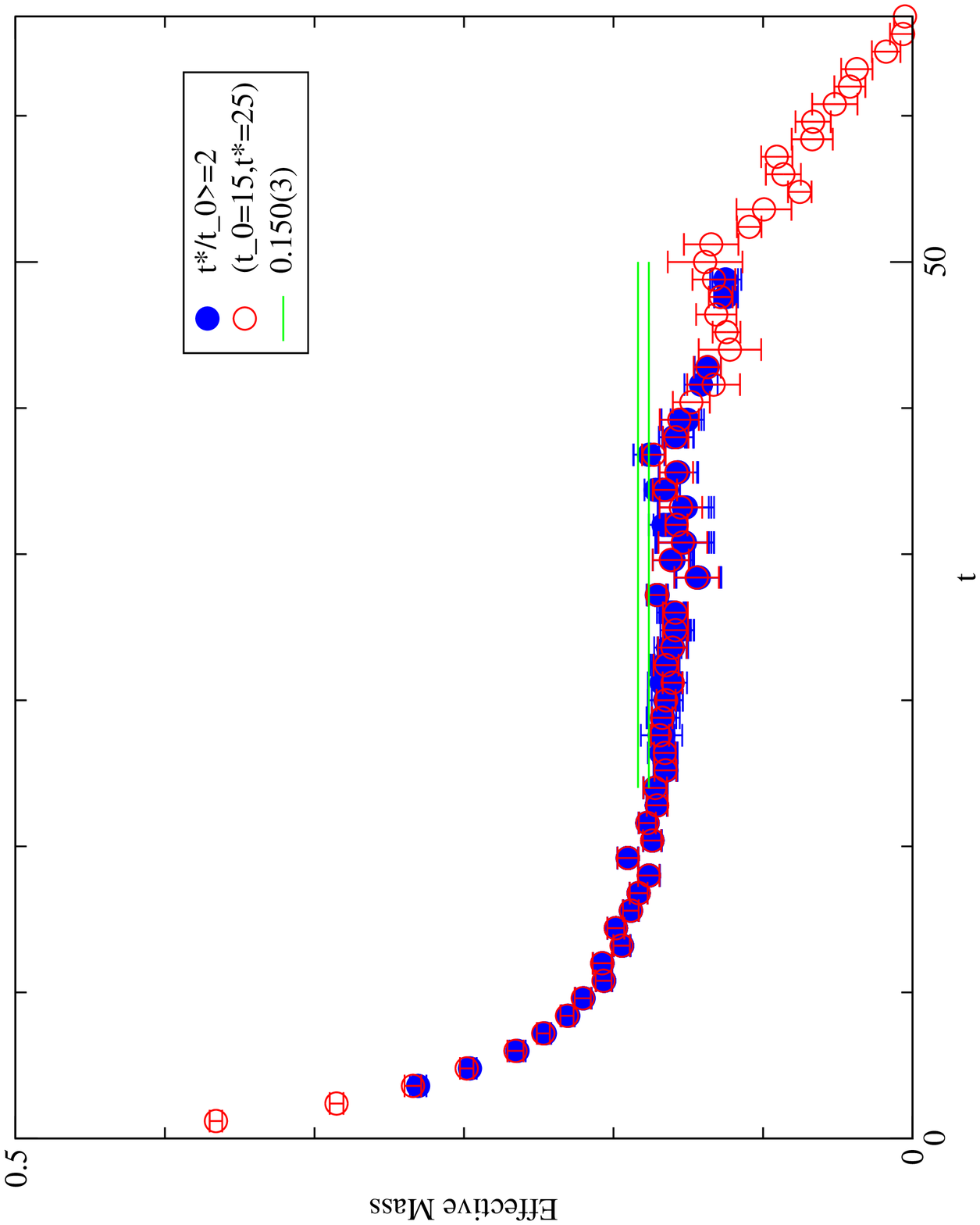} 
\caption{The optimized distilled pi-pi effective mass for the {\bf ground state} (${\bf t_0=15,t^*=25}$) on the $16^3\times128$ lattice with 64 eigenvectors. A single time-slice was used for the source operator and the standard definition of the effective mass was used with $\Delta t=1$.}
\end{center}
\label{fig:opt0_15}
\end{minipage}
\end{tabular}
\end{center}
\end{figure}

\begin{figure}[t]
\begin{center}
\begin{tabular}{cc}
\begin{minipage}{73mm}
\begin{center}
\ig[height=0.33\tht,width=0.28\tht,angle=270]{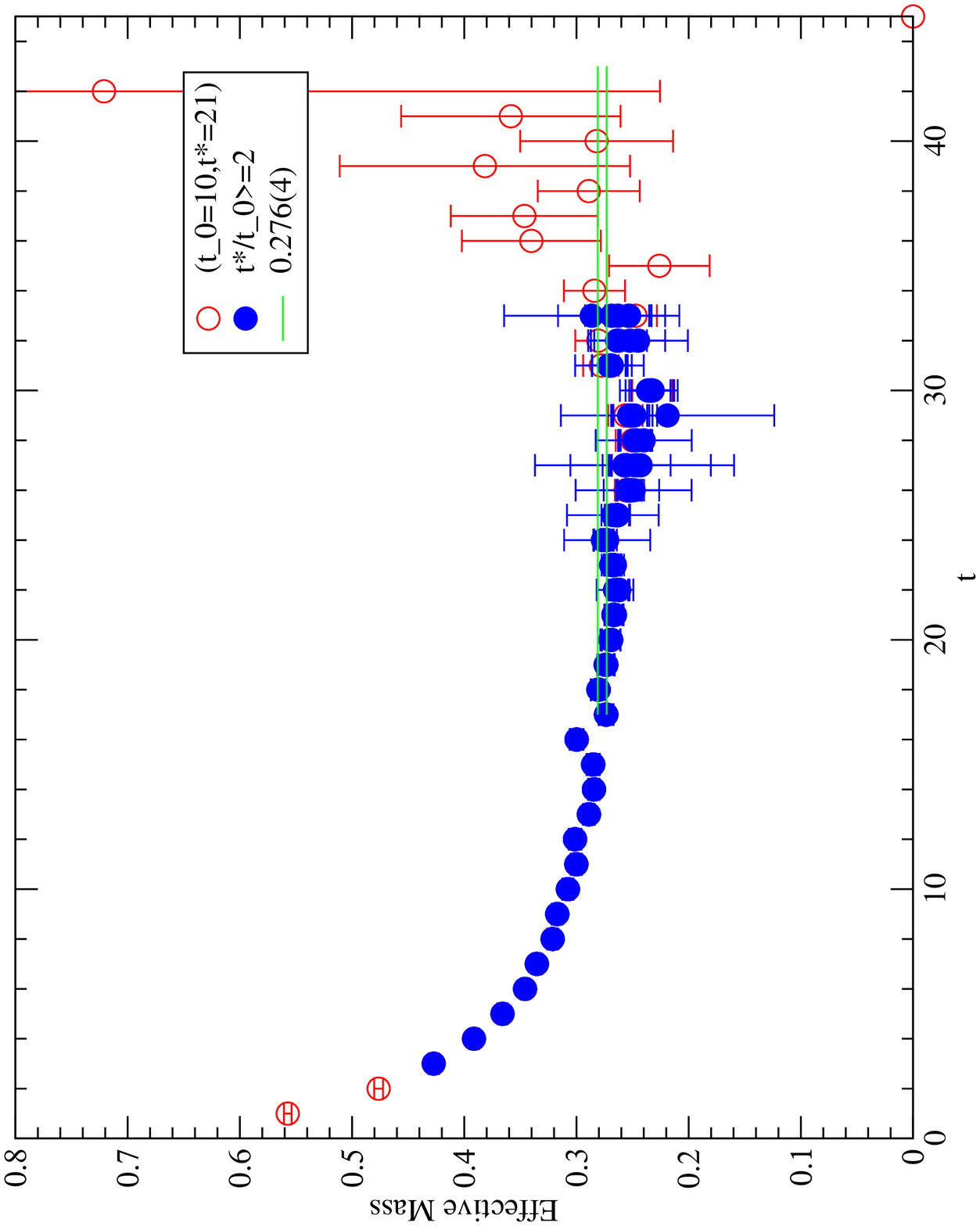} 
\caption{The optimized distilled pi-pi effective mass for the {\bf first excited state} (${\bf t_0=10,t^*=21}$) on the $16^3\times128$ lattice with 64 eigenvectors. A single time-slice was used for the source operator and the standard definition of the effective mass was used with $\Delta t=1$.}
\label{fig:opt1_10}
\end{center}
\end{minipage}&
\begin{minipage}{73mm}
\vspace{4mm}
\begin{center}
\ig[height=0.33\tht,width=0.28\tht,angle=270]{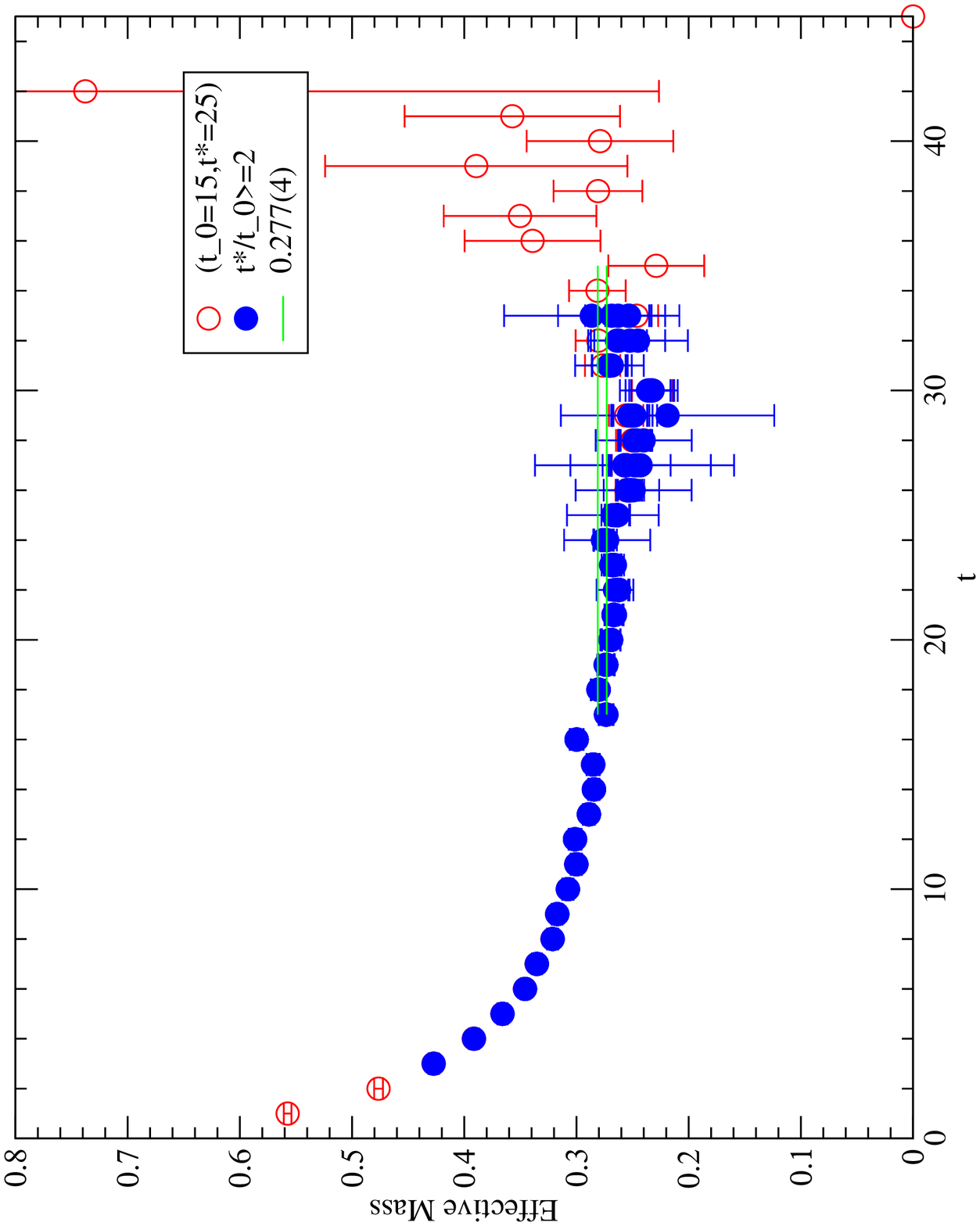} 
\caption{The optimized distilled pi-pi effective mass for the {\bf first excited state} (${\bf t_0=15,t^*=25}$) on the $16^3\times128$ lattice with 64 eigenvectors. A single time-slice was used for the source operator and the standard definition of the effective mass was used with $\Delta t=1$.}
\label{fig:opt1_15}
\end{center}
\end{minipage}
\end{tabular}
\end{center}
\end{figure}

The excited state contamination can be guaranteed to be from states higher than the $N^{th}$ level by choosing the ratio of $t^*$ to $t_0$ larger than two (Ref.~\cite{Blossier:2009kd}). Here we fix this ratio to two and plot the effective mass along side the fixed-coefficient method in order to check that the fixed coefficient signal has no contamination from the lower-lying excited states. We show examples for the ground state and first excited state to show that the systematics are under control (Fig.~\ref{fig:opt0_10}-\ref{fig:opt1_15}). 

We have also chosen two different sets of (${t^*,t_0}$) and computed their effective masses to show that the dependency there is also very small (Fig.~\ref{fig:opt0_10}-\ref{fig:opt1_15}). 

%%%%%%%%%%%%%%
%%%%%%%%%%%%%%

%%%%%%%%%%%%%%
%%%%%%%%%%%%%%

\subsection{Fitting}
Given the consistency of the various methods of diagonalization, we have chosen to fit the fixed coefficient, optimized correlators to compute the $\pi\pi$ energy levels. One can perform both a single exponential fit and two-exponential fits to the optimized correlators, but the two fits agree within statistical errors as can be seen in Fig.~5. We therefore extract the two-pion energy from single exponential fits whose fitting form was,
$$
C_{\pi\pi}(t)=A\left[e^{-(2M_{\pi}+\delta E)t}+e^{-(2M_{\pi}+\delta E)(T-t)}\right]+Be^{-M_\pi T}.
$$
Here, $A, B, E_{\pi\pi}$ and $M_\pi$ are the constants to be fit and $T=128$ is the time extent of the lattice. The scattering length is determined from L\"uscher's formula (Ref.~\cite{Luscher:1986pf}),
$$
a_t\delta E=-\frac{1}{\xi^2}\left(\frac{r_0}{a_s}\right)\frac{4\pi \tilde{a}_0}{(a_tM_\pi)(L/a_s)^3}\left\{1-2.837297\frac{\tilde{a}_0}{L/a_s}\left(\frac{r_0}{a_s}\right)+6.375183\frac{\tilde{a}_0^2}{(L/a_s)^2}\left(\frac{r_0}{a_s}\right)^2\right\}
$$
Preliminary results for the ratio $a_0/M_\pi$ extracted from uncorrelated fits in units of $\text{GeV}^{-2}$ is shown in Fig.~6. The statistical correlation between the single pion correlation functions and two-pion correlation function have not been taken into account in the fits, but the statistical error is expected to decrease when they are fit simultaneously as they are strongly correlated. Work is underway to fit the ratios simultaneously. 

%%%%%%%%%%%%%%
\begin{figure}[t]
\begin{center}
\begin{tabular}{cc}
\begin{minipage}{73mm}
\begin{center}
\ig[height=0.34\tht,angle=270]{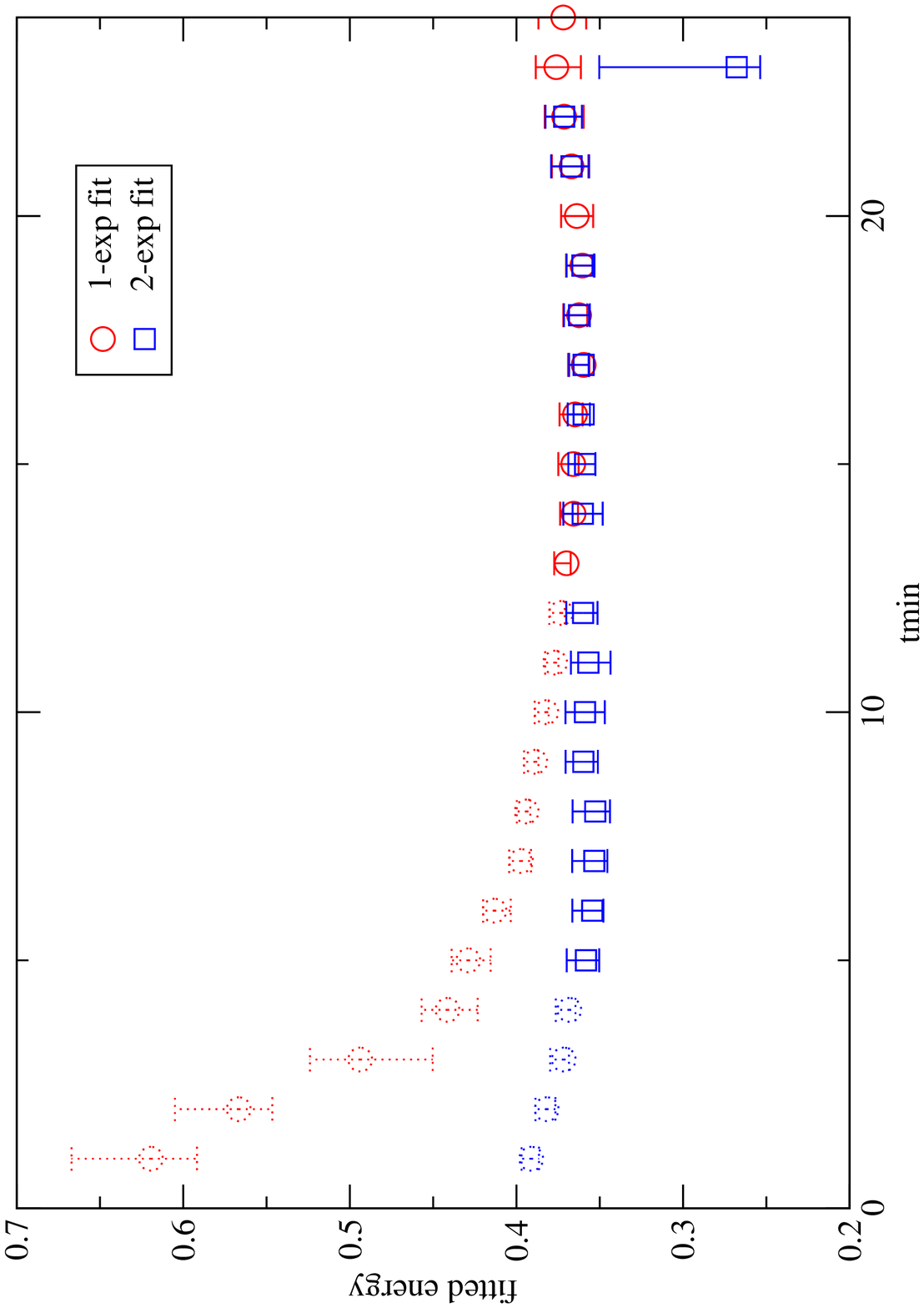} 
\caption{The $t_{min}$ plot of the fitted values for the pi-pi energies of the second excited state ($V=16^3$). The dashed fit values indicate a poor fit.} 
\end{center}
\label{fig:tmin}
\end{minipage}&
\begin{minipage}{73mm}
\begin{center}
\ig[height=0.23\tht]{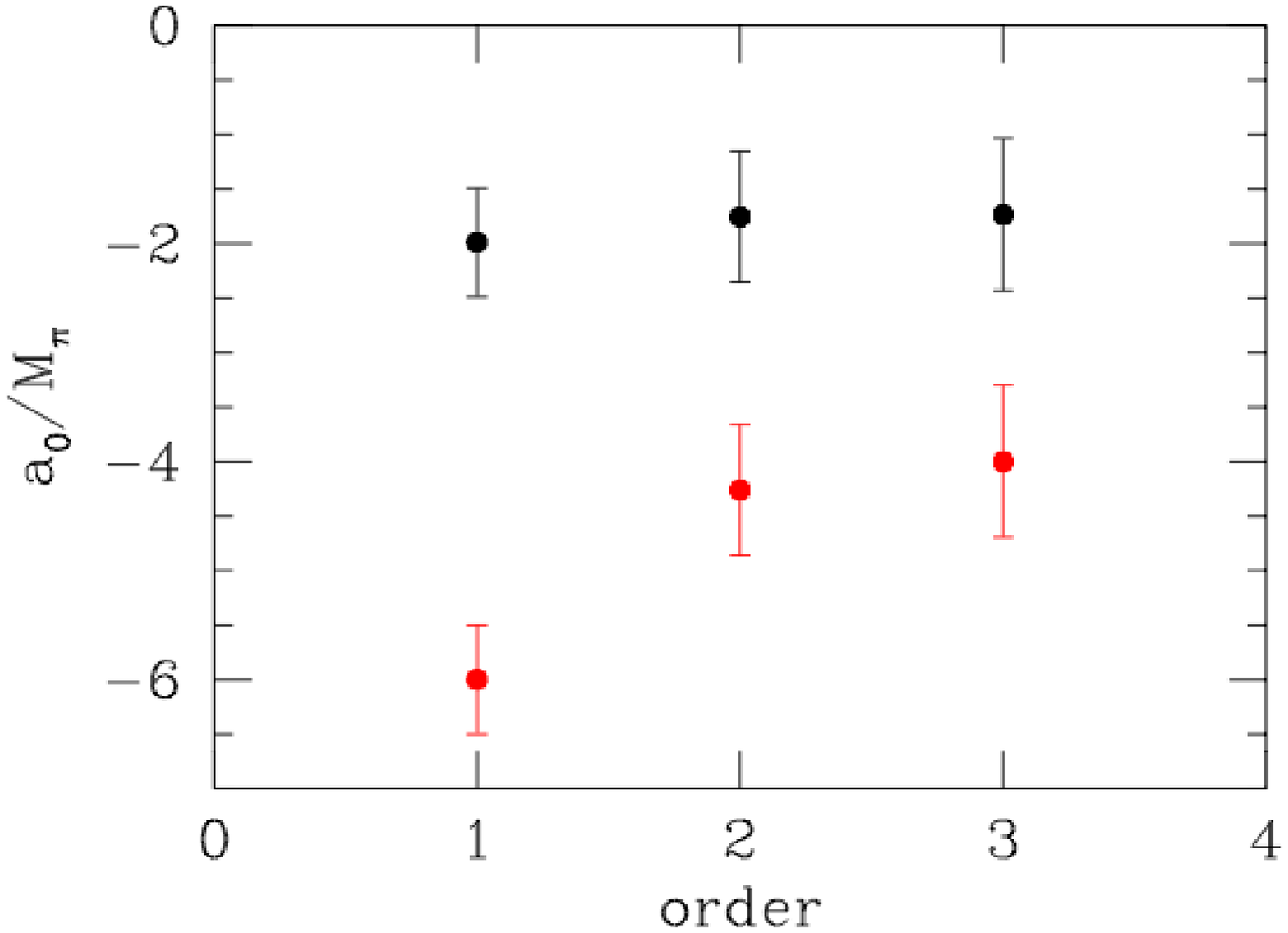}
\caption{Preliminary values for the scattering length in units of $GeV^{-2}$ extracted using the $1/L$ and $1/L^3$ formula (The $1/L^2$ value is shown for reference purposes only).}
\end{center}
\label{fig:Lcompare}
\end{minipage}
\end{tabular}
\end{center}
\end{figure}
%%%%%%%%%%%%%%

\section{Summary}
The simulation of the $I=2$ $\pi\pi$ scattering states using distilled quark propagators on $2+1$ anisotropic, dynamical lattices have been presented. A relatively light quark mass was used on two different volumes to check for finite size effects. The distilled propagators have allowed the extraction of the $4^{th}$ excited state to $10\%$ errors on $100$ configurations. Work is underway to incorporate the correlations between the single pion and the two-pion measurements to reduce the statistical errors in the determination of the scattering length and to determine the scattering phase shift. A modified version of the distillation procedure is also being investigated to reduce the computational/storage cost of the method in larger volumes and correlation functions which contain disconnected diagrams. 

\section*{Acknowledgements}
This work has been partially supported by National Science Foundation awards PHY-0704171. These calculations were performed using the Chroma software suite (Ref.~\cite{Edwards:2004sx}) on clusters at Jefferson Laboratory using time awarded under the SciDAC Initiative, clusters at University of the Pacific and Carnegie Mellon University.

\end{document}